%% file: latex_template_MedPhys_2021.tex
\documentclass[12pt,twoside]{article}   
\usepackage[super,sort&compress,comma]{natbib}
\usepackage{amsmath}
\usepackage{amsfonts}
\usepackage{booktabs}
\usepackage{fancyhdr}   

\usepackage[section]{placeins}   %

\usepackage{graphicx}

\makeatletter \renewcommand\@biblabel[1]{$^{#1}$} \makeatother
\newcommand{\cen}[1]{\begin{center} #1 \end{center}}

\usepackage{hyperref}
\hypersetup{ colorlinks,
    citecolor=blue,
    filecolor=blue,
    linkcolor=blue,
    urlcolor=blue
}

\usepackage{xcolor}

\definecolor{gray}{rgb}{0.6,0.6,0.6}
\definecolor{red}{rgb}{0.85,0,0}
\definecolor{green}{rgb}{0,0.85,0}
\definecolor{blue}{rgb}{0,0,0.85}
\definecolor{beige}{rgb}{0.92,0.87,0.78}
\usepackage[all]{hypcap}    

\begin{document}

\cen{\sf {\Large {\bfseries Accelerated respiratory-resolved 4D-MRI with separable spatio-temporal neural networks } \\  
\vspace*{10mm}
M. L. Terpstra$^{1,2,a}$, M. Maspero $^{1,2}$, J.J.C. Verhoeff$^{1}$, C.A.T. van den Berg$^{1, 2}$} \\
\vspace{5mm}
$^1$Departement of Radiotherapy, University Medical Center Utrecht, Utrecht, The Netherlands \\
$^2$Computational Imaging Group for MR Diagnostics \& Therapy, Center for Image Sciences, University Medical Center Utrecht, The Netherlands \\\
\vspace{5mm}\\
Version typeset \today\\
}

\pagenumbering{roman}
\setcounter{page}{1}
\pagestyle{plain}
a) Author to whom correspondence should be addressed. E-mail: \url{m.l.terpstra-5@umcutrecht.nl} \\

\clearpage     

\begin{abstract}
\noindent\textbf{Background}: Respiratory-resolved four-dimensional magnetic resonance imaging (4D-MRI) provides essential motion information for accurate radiation treatments of mobile tumors. However, obtaining high-quality 4D-MRI suffers from long acquisition and reconstruction times. \\
\textbf{Purpose}: To develop a deep learning architecture to quickly acquire and reconstruct high-quality 4D-MRI, enabling accurate motion quantification for MRI-guided radiotherapy.\\
\textbf{Methods}: A small convolutional neural network called MODEST is proposed to reconstruct 4D-MRI by performing a spatial and temporal decomposition, omitting the need for 4D convolutions to use all the spatio-temporal information present in 4D-MRI. 
This network is trained on undersampled 4D-MRI after respiratory binning to reconstruct high-quality 4D-MRI obtained by compressed sensing reconstruction. 
The network is trained, validated, and tested on 4D-MRI of 28 lung cancer patients acquired with a T1-weighted golden-angle radial stack-of-stars sequence. 
The 4D-MRI of 18, 5, and 5 patients were used for training, validation, and testing. 
Network performances are evaluated on image quality measured by the structural similarity index (SSIM) and motion consistency by comparing the position of the lung-liver interface on undersampled 4D-MRI before and after respiratory binning.
The network is compared to conventional architectures such as a U-Net, which has 30 times more trainable parameters.\\
\textbf{Results}: MODEST can reconstruct high-quality 4D-MRI with higher image quality than a U-Net, despite a thirty-fold reduction in trainable parameters. High-quality 4D-MRI can be obtained using MODEST in approximately 2.5 minutes, including acquisition, processing, and reconstruction.\\
\textbf{Conclusion}: High-quality accelerated 4D-MRI can be obtained using MODEST, which is particularly interesting for MRI-guided radiotherapy.

\end{abstract}



\setlength{\baselineskip}{0.7cm}      

\pagenumbering{arabic}
\setcounter{page}{1}
\pagestyle{fancy}
\section{Introduction}\label{sec1}
Respiratory motion poses a significant challenge in abdominal and thoracic imaging, causing large displacements in the liver\cite{https://doi.org/10.1118/1.4754658}, lung\cite{EKBERG199871}, kidney\cite{https://doi.org/10.1002/jmri.22418}, and pancreas\cite{FENG2009884}, introducing disruptive image artifacts that may preclude an accurate diagnosis\cite{Wood1985MRIA,NOTERDAEME2007273}. 
In radiation therapy, respiratory-induced motion can lead to sub-optimal treatment because it may influence the shape and position of tumors\cite{OZHASOGLU20021389,https://doi.org/10.1118/1.2349696}. Consequently, the target may receive a different dose than planned while delivering hazardous radiation to nearby healthy tissue and organs-at-risk\cite{BUSSELS200369}. 
In the past, respiratory-resolved imaging has been proposed to improve treatments, using imaging with high spatial resolution and accurate motion information to enable the definition of treatment margins that encompass the tumor displacement\cite{Vedam_2002,https://doi.org/10.1002/mrm.25753}. 
In particular, four-dimensional respiratory-resolved computed tomography (4D-CT) is the standard imaging modality in current clinical practice and is part of radiation treatment planning\cite{RIETZEL2006287}. However, 4D-CT can be affected by artifacts that negatively influence the treatment outcome and local control\cite{SENTKER2020229,https://doi.org/10.1118/1.2717404}.

Recently, magnetic resonance imaging (MRI) has been proposed as an alternative to CT for radiotherapy guidance, leveraging the superior soft-tissue contrast that facilitates accurate target identification and dose deposition.
With the clinical introduction of MRI-guided radiotherapy (MRIgRT)\cite{Raaymakers_2009,MUTIC2014196}, MRI acquired prior to treatment can be used to adapt the treatment plan to the daily anatomy, while fast MRI during treatment can be used to track the tumor position\cite{https://doi.org/10.1002/mp.15217,Huttinga_2020,Glitzner_2015,MENTEN2017139,https://doi.org/10.1118/1.4927252,KEALL2019228}.

In MRIgRT, respiratory-resolved four-dimensional MRI (4D-MRI) is used in the treatment planning phase to adapt the radiation treatment based on the quantified tumor motion \cite{PAULSON202072}. The 4D-MRI must be high-quality and quickly available to ensure treatment efficiency and patient comfort, i.e., acquired and reconstructed within five minutes\cite{Mickevicius_2017}. 
However, obtaining high-quality 4D-MRI remains challenging due to the limited acquisition speed of MRI.

A straightforward way to accelerate MRI is by undersampling the acquisition, violating the Shannon-Nyquist data sufficiency criterion\cite{6773024}, and introducing image artifacts that may preclude accurate motion quantification\cite{Mickevicius_2017}.
Several techniques have been proposed to reconstruct high-quality MRI from undersampled acquisitions, such as parallel imaging\cite{sense_99,https://doi.org/10.1002/mrm.10171}, simultaneous multi-slice acquisitions\cite{sms_01,Keijnemans_2021,https://doi.org/10.1002/mp.15802}, or compressed sensing\cite{https://doi.org/10.1002/mrm.21391}. 
Some algorithms have been specifically developed to reconstruct high-quality respiratory-resolved 4D-MRI by taking advantage of all spatio-temporal information in the images, such as XD-GRASP\cite{https://doi.org/10.1002/mrm.25665} or HDTV-MoCo\cite{https://doi.org/10.1002/mrm.26206}. 
However, these reconstruction algorithms have a large computational cost and can take from 15 minutes up to 8 hours\cite{PAULSON202072,https://doi.org/10.1002/mrm.26206}, which is insufficient in clinical practice as long treatment times are detrimental to patient comfort and treatment efficiency.

Recently, convolutional neural networks (CNNs) have been proposed as a data-driven alternative to classic iterative algorithms to reconstruct undersampled MRI quickly\cite{https://doi.org/10.1002/mrm.26977,8067520,10.1007/978-3-030-59713-9_7,https://doi.org/10.1002/mrm.27706,9048706}.
With CNNs, the time-consuming model training can be performed offline before treatment. Then, the trained model can be used for fast, online inference, achieving reconstruction quality on par or better than compressed sensing within tens of milliseconds for 2D imaging \cite{machado2022deep}.

Training such models requires large amounts of GPU memory to optimize the model parameters. 
As GPU memory is limited, training CNN-based reconstruction models is feasible for 2D and 3D MRI but challenging for 4D-MRI as these models require prohibitively costly four-dimensional convolutions to take advantage of the spatio-temporal information and obtain high image and motion quality.
Several approaches have been proposed to avoid using 4D convolutions, e.g., by performing slice-by-slice reconstruction or carefully using multiple views of the spatio-temporal data\cite{FREEDMAN2021209,Kuestner2020,8425639,8793147}.  
However, training such models to obtain high-quality 4D-MRI remains challenging due to the computational cost or requirement for large datasets.

We propose an unrolled model to reconstruct 4D-\textbf{M}RI using l\textbf{o}w-\textbf{d}im\textbf{e}nsional \textbf{s}ubne\textbf{t}works (MODEST), which exploits the spatio-temporal nature of 4D-MRI by separating the reconstruction problem into spatial and temporal components. 
Two independent subnetworks with few trainable parameters have been designed to learn these components without using 4D convolutional kernels. 
This allows the model to access the complete spatio-temporal information in 4D-MRI while maintaining low computational cost.

This work investigates the application of the proposed spatio-temporal decomposed network to accelerate the acquisition and reconstruction of undersampled 4D respiratory-resolved lung MRI, which is of particular interest for MRI-guided radiation treatments. 
The model is evaluated on reconstructed image quality and consistency of the respiratory motion compared to compressed sensing reconstructions. Moreover, MODEST is compared to standard deep learning architectures such as a U-Net. Finally, we estimate the minimum acquisition length for high-quality 4D-MRI with MODEST.

\section{Methods}
We considered two networks to reconstruct 4D-MRI: a baseline residual U-Net, and our newly proposed architecture. After patient data was collected and pre-processed, the model hyperparameters were optimized. Then, the U-Net and MODEST were trained. To investigate the impact of the model architecture rather than the number of trainable parameters, the optimized parameters of the U-Net were pruned to match MODEST. The three models (MODEST, the baseline U-Net, and pruned U-Net) were evaluated using undersampled 4D-MRI before and after respiratory binning. 
\subsection{Patient data collection and preparation}
Twenty-eight patients undergoing radiotherapy for lung cancer between February 2019 and February 2020 at the radiotherapy department were retrospectively included under the approval of the local medical ethical committee with protocol number 20-519/C.
The male/female ratio was 16/12, and the mean age was \(66\pm13\) years (range = 20-81). Patients affected by squamous cell carcinoma (11), adenoma \& adenocarcinoma (7), small cell/large cell carcinoma (4), neoplasm (1), thymoma (1), and a mix of other rare tumors (4) were included.

Free-breathing 3D golden-angle radial stack-of-stars (GA-SOS) \(T_1\)-weighted spoiled gradient echo MRI (TR/TE=3.2/1.3 ms, FA=8\(^\circ\), bandwidth=866Hz/px, resolution=\(2.13\times2.13\times3.5\) mm\(^3\), FOV=\(440\times440\times270\) mm\(^3\), feet-head slices) of the thorax were acquired for 7 min on a 1.5T MRI (MR-RT Philips Healthcare, Best, the Netherlands) during gadolinium injection (Gadovist, 0.1 ml/kg). The acquisition was fat-suppressed using spectral attenuated inversion recovery (SPAIR). 

Patients were scanned in the supine position using a 16-channel anterior and 12-channel posterior phased-array coil. In total, 1312 radial spokes per slice were acquired, corresponding to approximately four times oversampling compared to a fully-sampled volume, which requires \(206\cdot\pi/2\approx324\) spokes.
However, as the contrast agent was injected, the relative magnitude of the self-navigation signal changed over time.
To account for the contrast pickup phase, we discarded the first 200 spokes of every scan to prevent contrast mixing. 

For every patient, 4D-MRI was created based on a self-navigation signal by sorting k-space into ten respiratory-correlated bins for a final matrix size of \(V_x,V_y,n_{\text{slice}},n_{\text{phase}} = 206\times206\times77\times10\).
The self-navigation signal was obtained by performing a 1D Fourier transform of the center of k-space (i.e., \(k_0\)) along the slice direction and principal component analysis on the concatenated navigators\cite{https://doi.org/10.1002/mrm.25858,https://doi.org/10.1002/mrm.25665}.
Then, radial spokes were sorted into respiratory bins using a hybrid binning algorithm\cite{stemkens_hybrid} based on the phase and relative amplitude of the motion surrogate. 
For training purposes, undersampled 4D-MRI was obtained by undersampling the respiratory bins, i.e., "phase undersampling", ensuring motion consistency between the target reconstruction and undersampled MRI. The fully-sampled 4D-MRI contained \(n\) spokes per bin for every patient. Phase-undersampled 4D-MRI was created by retaining the first \(n/k\) spokes per bin, where \(k\in\mathbb{N}\) is the acceleration factor, for undersampling factors R\(_{\text{4D}}\) = 1, 2, and 4. This corresponded to a true undersampling factor R\(_{\text{Nyquist}}\) of approximately 3.7, 7.4, and 14.8 per respiratory phase, respectively.
After sorting, k-space was density-compensated using a Ram-Lak filter, interpolated onto a twice-oversampled Cartesian grid using a \(3\times3\) Kaiser-Bessel kernel, and transformed to image-space using a non-uniform fast Fourier transform (NUFFT)\cite{1166689,knoll2014gpunufft} with a weighted coil combination. Coil sensitivity maps were estimated using ESPiRIT\cite{https://doi.org/10.1002/mrm.24751}. 
The patients were randomly split into a train (18), validation (5), and test (5).
The training target was generated by performing an XD-GRASP reconstruction of the fully-sampled 4D-MRI using temporal total variation, using a regularization weight \(\lambda=0.03\)\cite{Lustig2007,https://doi.org/10.1002/mrm.25665}.

To match the effect of a shorter acquisition time, we have also created undersampled 4D-MRI by removing spokes prior to respiratory binning and discarding the final \(j\) sampled spokes, with \(j \in \{100, 200, \ldots, 1000\}\), i.e., "free-breathing undersampling".
These reconstructions were used to estimate the maximum achievable undersampling factor in a clinical setting, comparing the motion consistency of the free-breathing undersampled 4D-MRI to the fully-sampled reconstruction. 
We selected the maximum value of \(j\) where the zero-filled reconstruction has a mean EPE \(<1\) mm and the mean SSIM of MODEST was \(> 0.85\).

\subsection{Model architectures}
\begin{figure*}
    \centering
    \includegraphics[width=\linewidth]{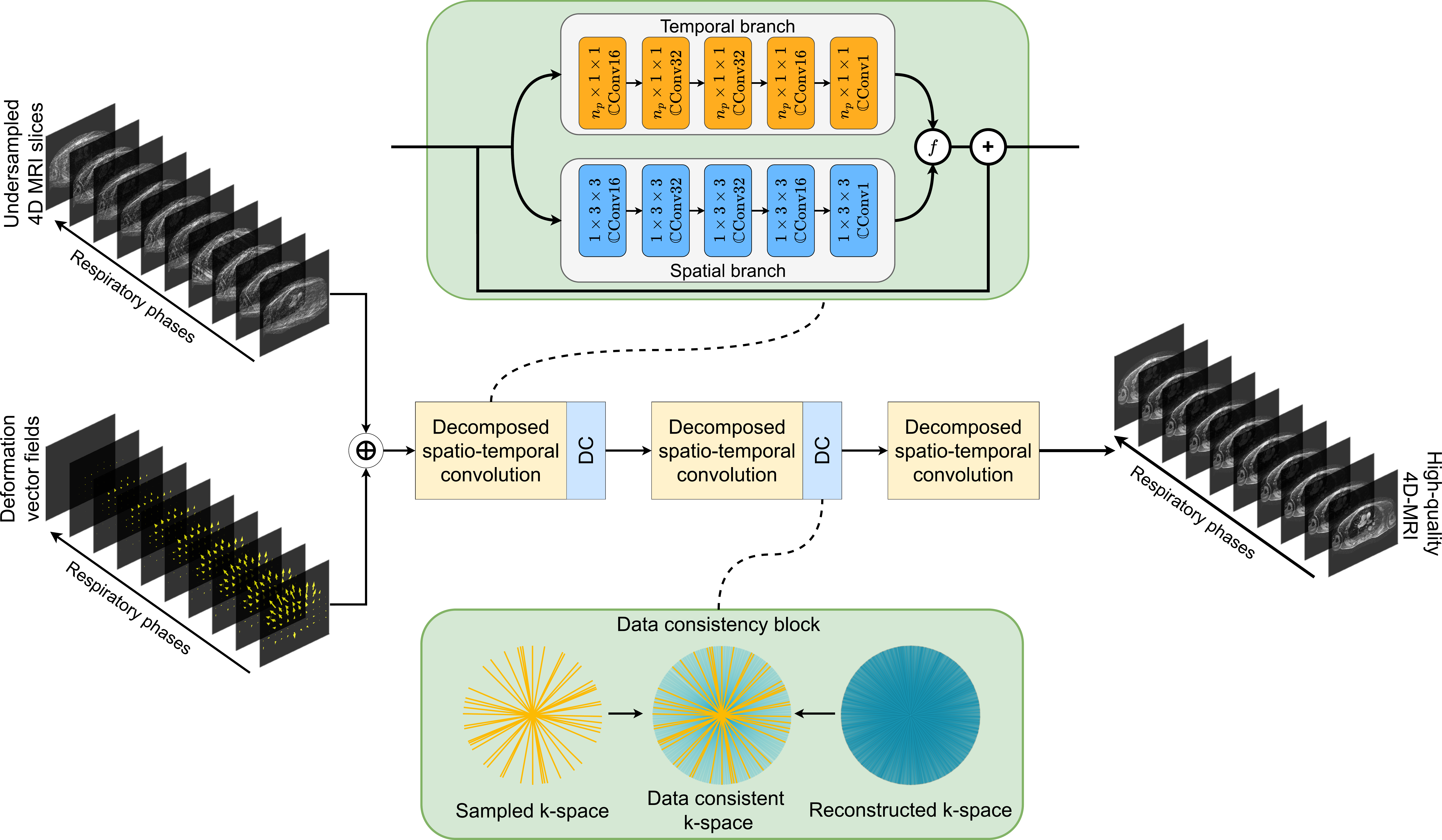}
    \caption{\textbf{Illustration of the proposed MODEST model}. The unrolled model reconstructs undersampled 4D-MRI into high-quality 4D-MRI. The undersampled, zero-filled 4D-MRI and deformation vector fields derived from the undersampled 4D-MRI are concatenated and enter a decomposed spatio-temporal convolution block with 104,260 parameters. The spatio-temporal convolution block performs low-dimensional convolution over the spatial domain (blue) and the temporal domain (orange), recombining into a 4D-MRI using a combination function \(f\). After every iteration of the unrolled model, data consistency is enforced on the reconstructed radial k-space using the sampled radial k-space using \autoref{eq:dc}. }
    \label{fig:network_arch}
\end{figure*}

We propose MODEST, which uses two subnetworks to learn the spatial and temporal features\footnote{Code available at \url{https://gitlab.com/computational-imaging-lab/modest}}. We trained a network to reconstruct 4D-MRI on a per slice basis rather than per volume to reduce memory usage, which allowed using 2D convolutions. The model input consisted of the zero-filled undersampled 4D-MRI and deformation vector fields (DVFs) computed on zero-filled, undersampled 4D-MRI, registering the exhale phase to every other respiratory phase.
The DVFs were obtained using a deep learning model\cite{Terpstra_2020}. They were added as additional input as we hypothesize that adding DVFs improves the reconstruction performance as they provide additional spatial information when considering the respiratory phase dimension.
To reconstruct a \(V_x\times V_y\times n_{\text{phase}}\) volume, the subnetwork learning the spatial component \(\hat{\Xi}\) was implemented using \(k\times k\times 1\) convolution kernels, while the network learning the temporal component \(\hat{\Psi}\) was implemented using \(1\times1\times n_{\text{phase}}\) convolutions. 
Both subnetworks used five convolutional layers and a cardioid non-linear activation function\cite{8297024}. The model hyperparameters and architecture were optimized using Bayesian optimization. Details for this optimization are provided in Supplementary Document 1.
An estimate of the 4D-MRI is then obtained as \(f(\hat{\Xi}, \hat{\Psi})\), using some combination function \(f\), which was chosen as the point-wise multiplication operator. 
We implemented the model to perform an unrolled optimization using three iterations.
Data consistency was enforced between the reconstructed image and the sampled k-space after every iteration except the final iteration by computing 
\begin{equation}
    \label{eq:dc}
    x^{t+1} = x^t - \eta\mathcal{F}^{-1}\left(\mathcal{F}\left(x^t\right)- \mathbf{y}\right) + \operatorname{M}_t\left(x^t\right),
\end{equation}
where \(t\) is the iteration, \(x^t\) is the image at iteration \(t\), \(\mathbf{y}\) is the measured, undersampled radial k-space, \(\mathcal{F}\) is the multi-coil non-uniform Fourier transform operator, \(\eta\) is a learned parameter, and \(\operatorname{M}_t\) is the deep learning model for iteration \(t\) of the unrolled model.
The model architecture is illustrated in \autoref{fig:network_arch} and had 312,782 trainable parameters.
To investigate the impact of data consistency and adding DVFs as model input, we have trained four variants of MODEST: a variant that only uses the zero-filled 4D-MRI, a variant that uses 4D-MRI and DVFs, a variant that uses 4D-MRI and data consistency, and a variant that uses 4D-MRI, DVFs, and data consistency.

MODEST was compared to a baseline residual U-Net\cite{10.1007/978-3-030-12029-0_40,the_monai_consortium_2020_4323059} that reconstructs 4D-MRI from the undersampled images, where every residual unit consisted of a 3D convolution layer, followed by a PReLU non-linear activation, instance normalization, and a residual connection. 
The residual U-Net consisted of four resolution levels and five residual units per resolution level. Depending on the resolution level, the residual unit's convolution layers learned 32, 64, 128, and 256 filters. 
The residual U-Net had 11,793,289 trainable parameters.
The model architecture and hyperparameters were found after a Bayesian hyperparameter search. Details for this optimization are provided in Supplementary Document 1. 

\subsubsection{Training and evaluation}
Both the residual U-Net and MODEST were implemented using PyTorch 1.10. The data consistency operator was implemented using TorchKbNUFFT 1.3.0\cite{muckley:20:tah}.
The U-Net and MODEST with optimized hyperparameters and architectures were trained on phase-undersampled MRI to reconstruct XD-GRASP 4D-MRI from zero-filled undersampled 4D-MRI. Both models were trained using 20,000 randomly-sampled batches of zero-filled 4D-MRI with undersampling factors R\(_{\text{4D}}\) = 1, 2, and 4 to minimize the \(\perp+\text{SSIM}\)-loss\cite{perploss}. In total, \(18\text{ patients} \cdot 77\text{ slices}\cdot 3\text{ undersampling factors} = 4185\) samples were used for training, and 1155 samples were used for testing and validation, respectively.
MODEST was trained using a batch size of 7 with the AdamW optimizer using a learning rate of \(10^{-3}\) and \(10^{-4}\) weight decay.
The baseline residual U-Net was trained using a batch size of 3 using the AdamW optimizer using a learning rate of \(10^{-3}\) and \(10^{-4}\) weight decay.
To investigate the impact of the model architecture rather than the number of trainable parameters, we performed iterative pruning of the trained U-Net model (Pruned U-Net), matching the number of parameters of MODEST\cite{DBLP:journals/corr/JinYFY16}.

The model reconstructions were evaluated on image quality, sharpness, motion quality, and processing time. The image quality was measured by the average SSIM and the normalized root-mean-square error (NRMSE) over the respiratory phases between the model reconstruction and the XD-GRASP reconstruction. The NRMSE was computed as \(\operatorname{NRMSE}\left(I_\text{est}, I_\text{target}\right) = \sqrt{1/M \sum\left(I_\text{est}- I_\text{target}\right)^2} / \overline{\vert I_\text{target}\vert}\), where \(M\) is the number of voxels and \(\overline{\vert I_\text{target}\vert}\) is the mean absolute value of \(I_\text{target}\) within the anatomy\cite{https://doi.org/10.1002/mp.15514}. 
The motion estimation quality was quantified in two ways:
\begin{enumerate}
    \item DVFs based on XD-GRASP reconstructions and the deep learning reconstructions were estimated using a neural network trained on undersampled MRI\cite{Terpstra_2020}, registering the first respiratory phase (exhale) to every other respiratory phase. The motion error was then quantified as the mean end-point error (EPE).
    \item The position of the hepatic dome in the reconstruction was compared to the hepatic dome position in the ground-truth XD-GRASP reconstruction. The hepatic dome position was manually extracted by computing the median intensity along the AP direction and thresholding the gradient image\cite{VANDELINDT2018875}. Then, the liver position was estimated for every dynamic as the mean of the binary thresholded image along the LR direction within a manually delineated region, ensuring a similar delineation volume among the patients in the test set. The hepatic dome position was normalized by subtracting the position of the hepatic dome in the free-breathing zero-filled acquisition. Finally, the error was determined as the absolute error between the hepatic dome of XD-GRASP reconstructions and MODEST.  
\end{enumerate}
The image sharpness was evaluated over all the 4D-MRI phases by computing the variance of one 3D respiratory phase after convolution with a 3D Laplacian kernel\cite{903548}. The final sharpness was estimated as the mean variance over all respiratory phases. Sharper images have a higher variance.

The metrics' statistical significance (\(p < 0.05\)) was established using a paired t-test, comparing MODEST to the U-Net and parameter-pruned U-Net.
\section{Results}
\begin{figure*}
    \centering
    \includegraphics[width=\linewidth]{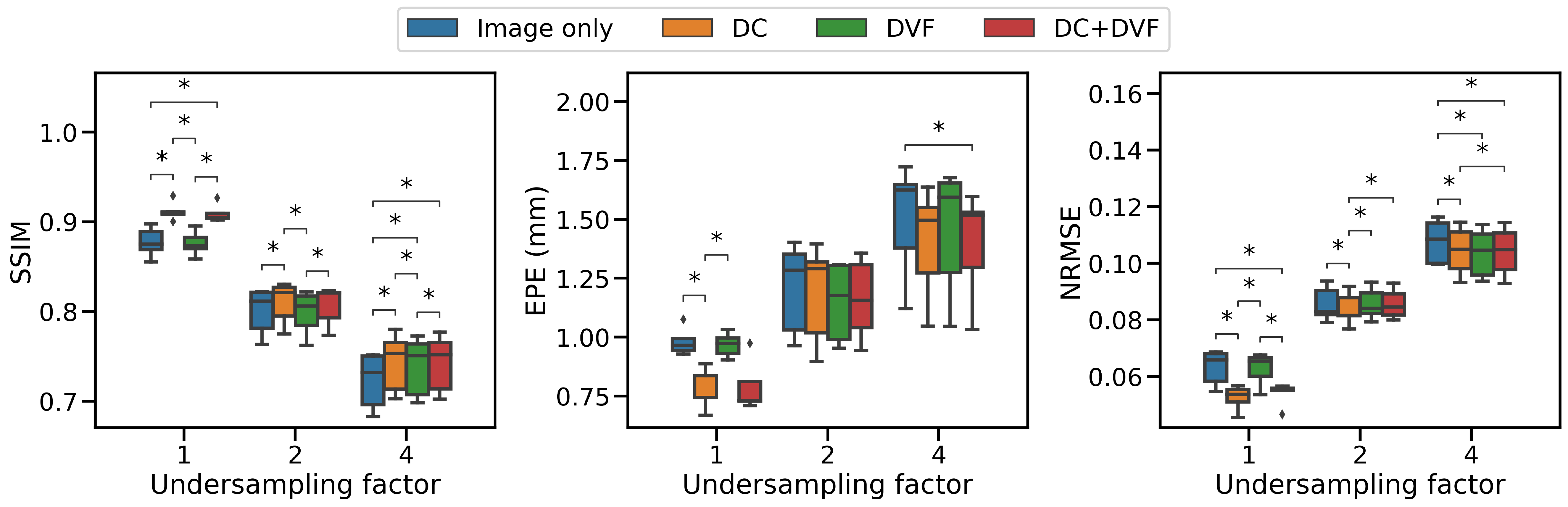}
    \caption{\textbf{Impact of data consistency and DVFs.} Four models were compared on the SSIM, registration error, and NRMSE in the foreground for the reconstructed 4D-MRI. The models used only the 4D-MRI, 4D-MRI and data consistency, 4D-MRI and DVFs, or all information to reconstruct the data. A star indicates a statistically significant result \(p < 0.05\).}
    \label{fig:comparison}
\end{figure*}

Based on the model architecture and hyperparameter search, we found that adding non-Cartesian data consistency and motion information increased the reconstruction quality, as shown in \autoref{fig:comparison}. Using data consistency increased the validation SSIM from \(0.88\pm0.04\) to \(0.90\pm0.04\) (\(p=10^{-6}\)), while adding DVFs did not significantly improve the SSIM compared to image-only reconstruction or in addition to using data consistency. However, using DVFs decreased the mean EPE from \(1.23\pm0.28\) mm to \(1.18\pm0.27\) mm (\(p=0.0008\)) and the NRMSE from \(0.086\pm0.02\) to \(0.084\pm0.18\) (\(p=0.0009\)), indicating increased motion consistency. Therefore, we opted to use data consistency and DVFs for MODEST. 

\subsection{4D-MRI reconstruction}
Phase-undersampled zero-filled reconstructions were created using a NUFFT in approximately 5 seconds, while the XD-GRASP reconstruction took about one hour.
MODEST took 15 seconds to process the zero-filled reconstructions on an NVIDIA V100 GPU, while the U-Net took approximately 30 seconds to reconstruct the 4D-MRI. The parameter-pruned U-Net took about 25 seconds to perform a reconstruction.

\begin{figure}
    \centering
    \includegraphics[width=0.9\columnwidth,trim={1mm, 1mm, 1mm, 1mm},clip]{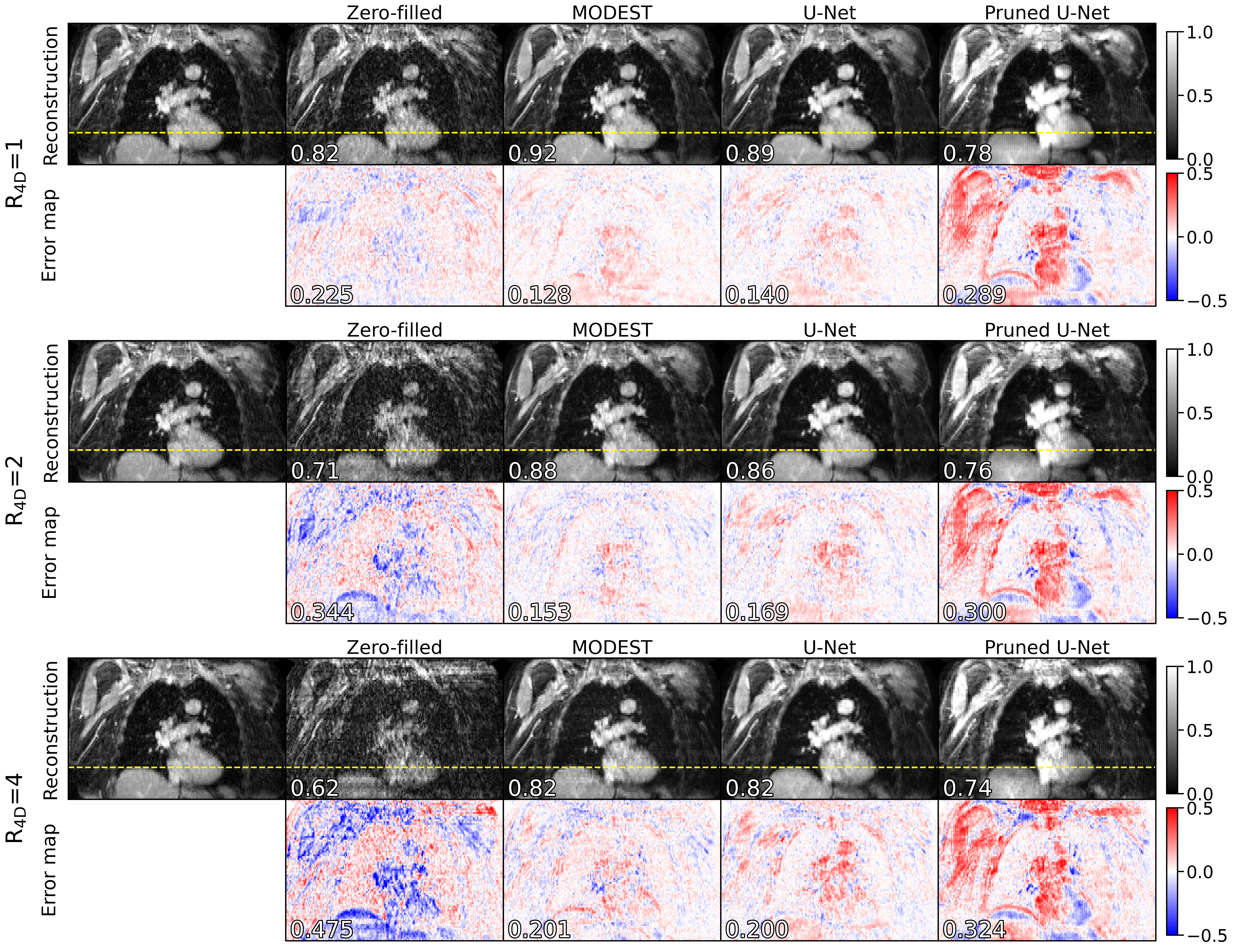}
    \caption{\textbf{Example reconstructions.} 4D-MRI was acquired of a female, 81 years old, affected by adenoma (T2N3M1). Reconstructions of phase-undersampled 4D-MRI inhaling by zero-filling, MODEST, the U-Net, and the parameter-pruned U-Net are shown for several undersampling factors and are compared to the XD-GRASP reconstruction. The top row shows the magnitude reconstructions and the SSIM, while the bottom row shows the NRMSE map and the mean NRMSE value for the 4D reconstruction. In \autoref{fig:quant_compare}, a quantitative evaluation for the entire test set is shown.}
    \label{fig:retro_recon}
\end{figure}

In the example of phase-undersampled 4D-MRI at \(R_{\text{4D}}=1\) in the test set (\autoref{fig:retro_recon}), MODEST produced reconstructions with an SSIM of 0.92 over the entire 4D volume, considering XD-GRASP as reference. This has significantly higher quality than the zero-filled reconstruction, which already shows undersampling artifacts and an SSIM of 0.82 (\(p = 0.0017\)).
Despite having over thirty times fewer trainable parameters, MODEST also produces higher image quality for the considered subject than the U-Net. Compensating for the increase in parameters of the U-Net, the pruned U-Net reconstructs 4D-MRI with low image and low motion consistency, as identified by the hepatic dome position.
At \(R_{\text{4D}}=4\), MODEST and U-Net showed comparable performance. However, the reconstructions by the U-Net seemed to suffer more from temporal blurring, as observable in the error maps of \autoref{fig:retro_recon}.
Videos of phase-undersampled reconstructions are provided for \(R_\text{4D}=1\text{, }2\text{, and }4\) in Supplementary Videos V1, V2, and V3, respectively. In these videos, it can be observed that MODEST and U-Net display similar image quality. However, in Supplementary Video V3, it can be seen that the U-Net reconstruction suffers from significantly reduced respiratory amplitude at the anterior chest wall, while MODEST shows better motion consistency.

\begin{figure*}
    \centering
    \includegraphics[width=0.7\linewidth]{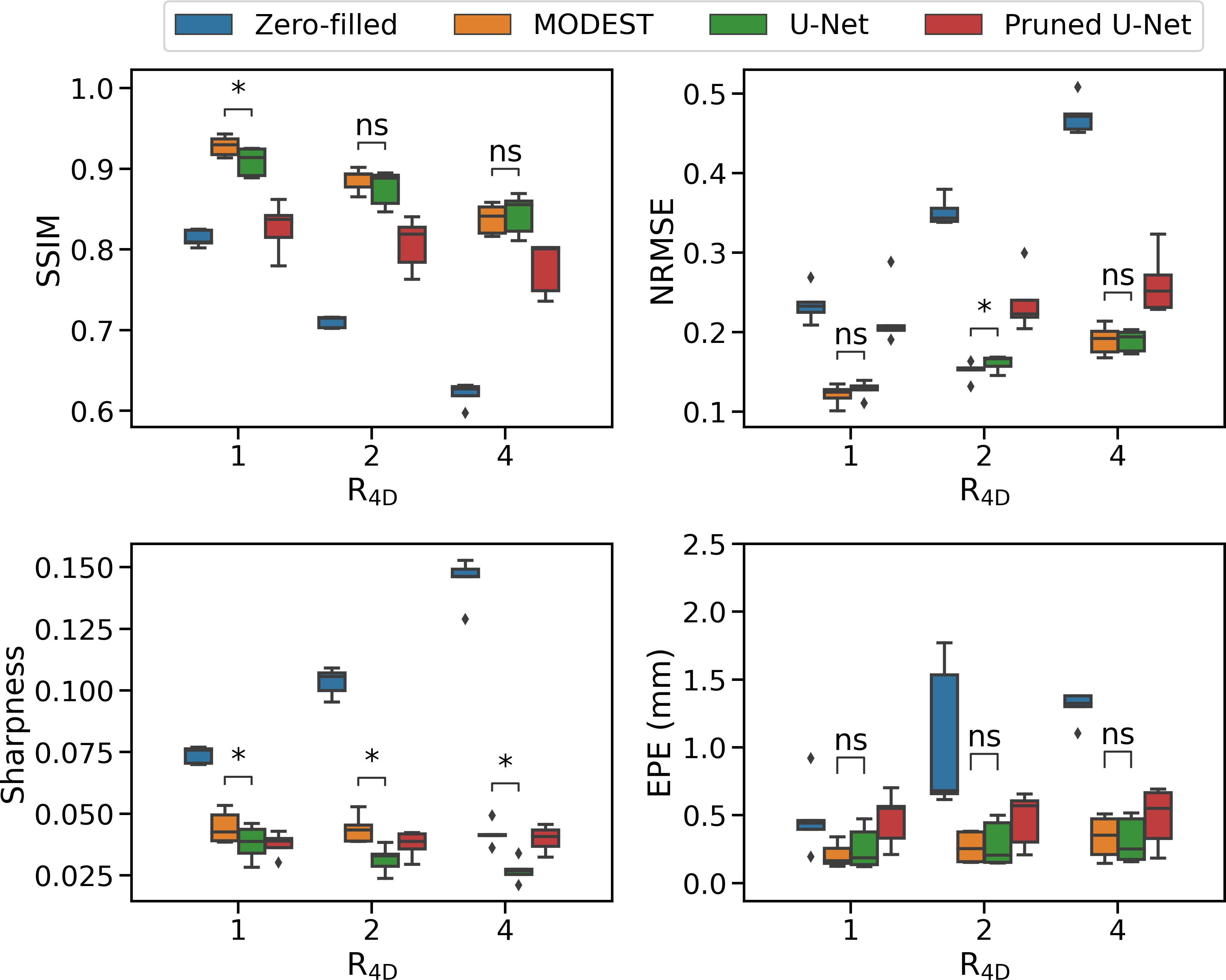}
    \caption{\textbf{Quantitative comparison.} All reconstruction methods are evaluated on the test set compared to the XD-GRASP reconstruction based on image similarity, measured by the SSIM and NRMSE, and motion similarity, measured by the EPE. All deep learning models perform significantly better than the zero-filled reconstruction, but MODEST outperforms the U-Net models based on image sharpness and NRMSE. A star indicates the t-test resulted in statistically significant differences with \(p < 0.05\).}
    \label{fig:quant_compare}
\end{figure*}

The U-Net and MODEST outperformed the zero-filled reconstruction based on the SSIM and EPE metrics (\(p = 10^{-9}\)), as visible in the quantitative evaluation in \autoref{fig:quant_compare}. However, no statistically significant difference was found between the U-Net and proposed architecture, except for the SSIM at \(R_{\text{4D}}=1\). Both models outperformed the parameter pruned U-Net for the SSIM metric (\(p = 10^{-8}\)). For the NRMSE metric, MODEST outperformed the U-Net, parameter pruned U-Net, and zero-filled reconstruction (\(p = 10^{-7}\)). 
MODEST showed sharper reconstructions for all under-sampling factors than the U-Net (\(p = 10^{-7}\)).

\begin{figure}
    \centering
    \includegraphics[width=0.99\linewidth]{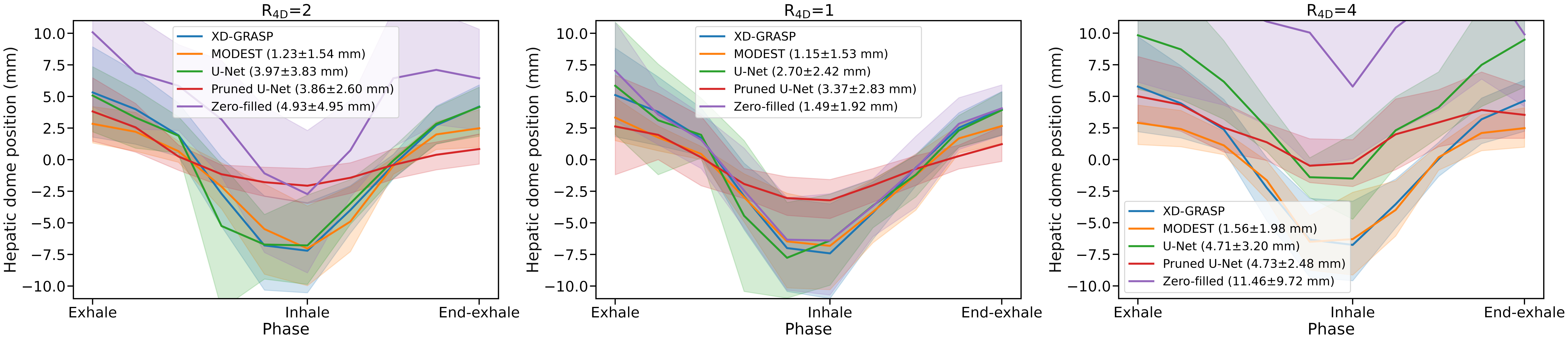}
    \caption{\textbf{Hepatic dome analysis.} MODEST closely follows the XD-GRASP reconstruction, especially at inhale. At high undersampling factors, MODEST is able to reconstruct motion-consistent 4D-MRI as measured by the hepatic dome, while the other reconstruction methods show significant errors.}
    \label{fig:liver_dome}
\end{figure}

Using MODEST led to reconstructions with increased motion consistency, as found by the increased correspondence of the hepatic dome position, as presented in \autoref{fig:liver_dome}. At \(R_{\text{4D}}=4\), the proposed architecture accurately tracked the hepatic dome position within \(1.56\pm1.98\) mm compared to the XD-GRASP reconstruction versus \(4.73\pm2.48\) for the U-Net. We observed that MODEST performed worse at exhale than inhale. However, the mean hepatic dome error was approximately 1.2 mm, significantly smaller than the voxel size of 3.5 mm in the feet-head direction.

\begin{figure}
    \centering
    \includegraphics[width=0.9\columnwidth]{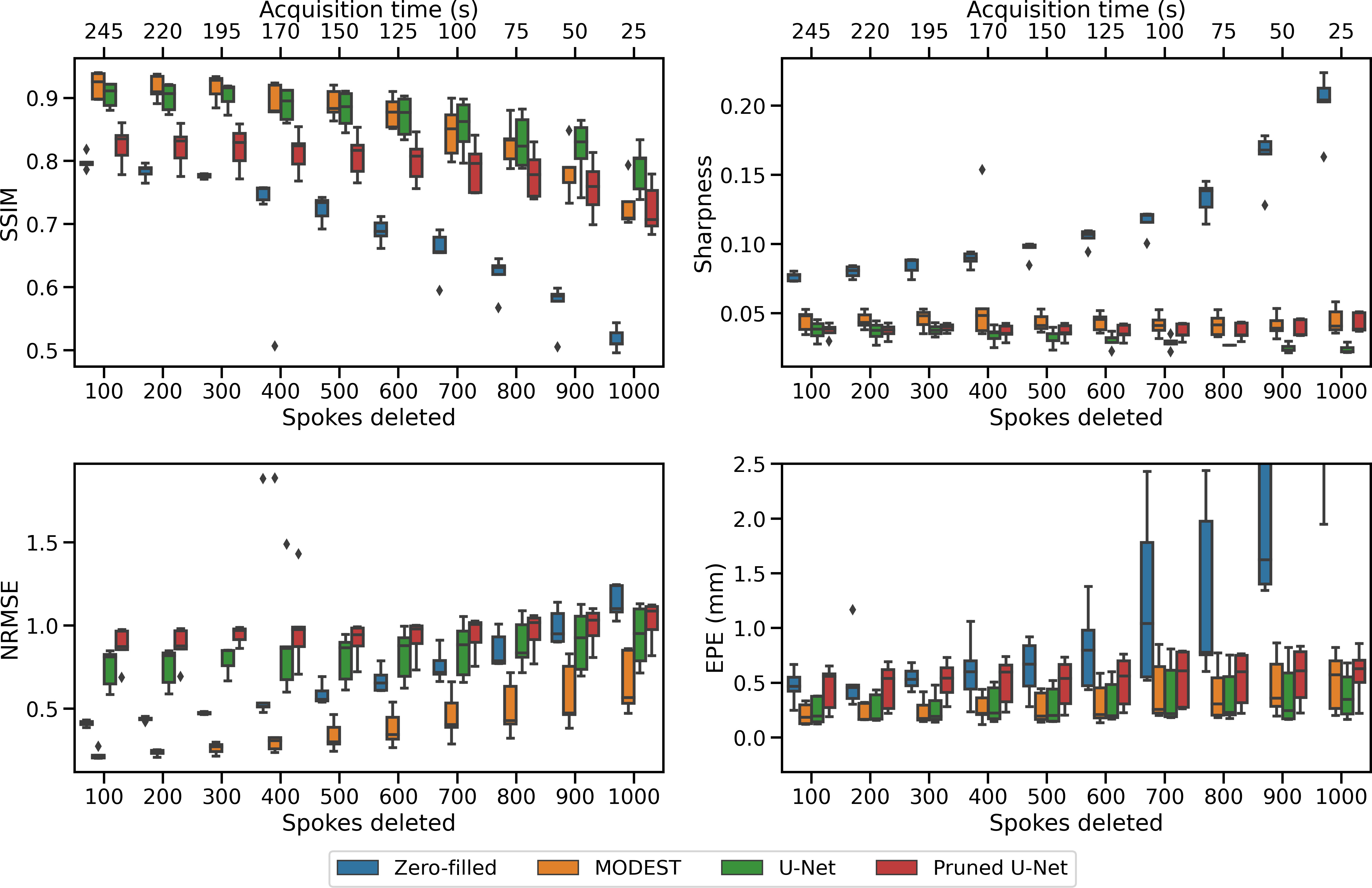}
    \caption{\textbf{Impact of free-breathing undersampling.} The impact of free-breathing undersampling was evaluated by continually removing \(n\) spokes from the acquisition and compared to the fully-sampled XD-GRASP reconstruction using the SSIM, EPE, and NRMSE metrics. As the increased significantly beyond removing 600 spokes, the minimum acquisition length was determined as 500 spokes. The approximate acquisition time is shown on top.}
    \label{fig:prospective_quant}
\end{figure}

Retaining fewer spokes for the free-breathing undersampled 4D-MRI decreased model performance due to an increased undersampling factor and increased intra-bin variability of the motion, as presented in \autoref{fig:prospective_quant}. 
The sharpness of the U-Net reconstruction decreased due to temporal blurring as the undersampling factor increased. In contrast, the sharpness of MODEST reconstruction is more stable. Based on the criterion that the shortest acquisition needed to have an EPE \(< 1\) mm for the zero-filled reconstruction and an SSIM \(>\) 0.85 for the MODEST reconstruction, using the first 500 spokes is the shortest free-breathing acquisition that allowed reconstructing high-quality 4D-MRI using MODEST, corresponding to an acquisition time of approximately two minutes. 

\begin{figure}
    \centering
    \includegraphics[width=0.98\columnwidth]{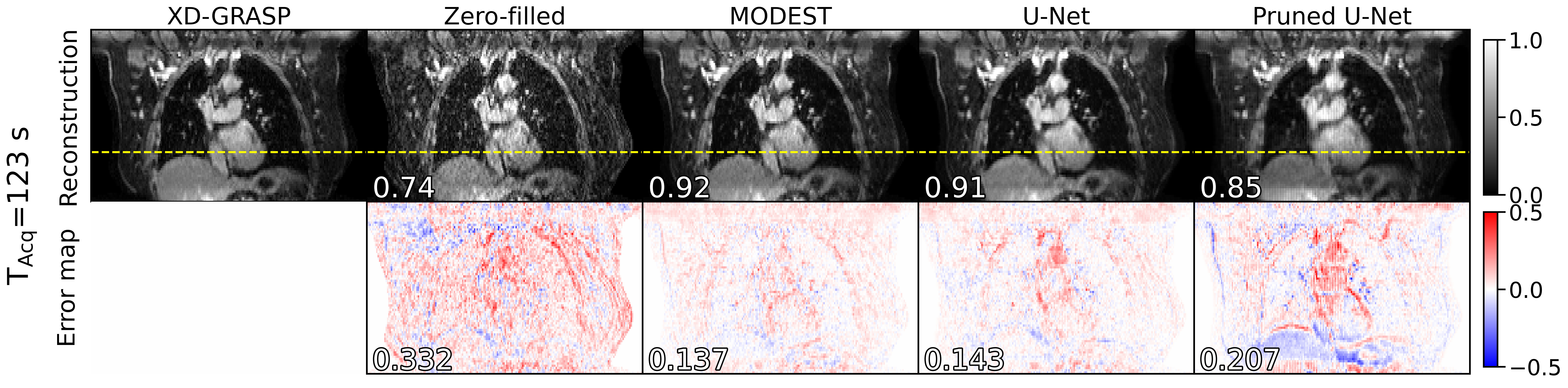}
    \caption{\textbf{Example free-breathing undersampled reconstructions.} 4D-MRI was acquired of a female, 71 years old, affected by squamous cell carcinoma (T2N1M0). Reconstructed 4D-MRI by MODEST, a U-Net, and the parameter-pruned U-Net are shown using an acquisition of 500 spokes (approx. 2 minutes) and are compared to the target XD-GRASP reconstruction. The top row shows the magnitude reconstructions and the SSIM, while the bottom row shows the NRMSE map and the mean NRMSE value for the 4D reconstruction.}
    \label{fig:prospective_recon}
\end{figure}

An example reconstruction for this acquisition is shown in \autoref{fig:prospective_recon}. Here, it can be seen that MODEST can reconstruct 4D-MRI with high quality with a mean SSIM of 0.92 and a mean NRMSE of 0.137 for this patient, which is of higher quality than the U-Net and pruned U-Net reconstruction. This model also shows good motion correspondence, as indicated by the alignment of the hepatic dome position. 
The quantitative results for the test set are presented in \autoref{tab:quant_600_tab}, showing that MODEST can achieve superior reconstructions compared to the U-Net and pruned U-Net, with an NRMSE of \(0.383\pm0.11\) versus \(0.824\pm0.16\) and \(0.920\pm0.11\), respectively.
A video of free-breathing undersampled reconstructions is provided in Supplementary Videos V4.

\input{table}
\section{Discussion}
In this work, we have proposed an architecture called MODEST for efficient 4D-MRI reconstruction by splitting the model into spatial and temporal components. We designed a model that exploits all spatio-temporal information of 4D-MRI using only low-dimensional convolution layers.
High-quality 4D-MRI was obtained using this model from highly undersampled acquisitions in only 25 seconds and outperforms an optimized residual U-Net, despite having 3\% of its trainable parameters.
We have shown that the model can accurately reconstruct 4D-MRI from shortened acquisitions for up to two minutes while maintaining high image quality (SSIM of \(0.877\pm0.025\)) and motion-consistency with the fully-sampled 4D-MRI.
These properties have some advantages over other models: models with few trainable parameters are less likely to overfit than larger models and have the potential to generalize better on unseen data due to less parameter variance \cite{6796174}. Moreover, small models typically require fewer training samples converge\cite{10.1093/bib/bbx044}, which is particularly interesting for MRI, as large datasets are difficult to acquire. 
 
Our hyper-parameter optimization and model architecture search found that performing data consistency improved image quality, and adding motion information increased the reconstructed image quality. These findings are in line with previously published literature\cite{9703109}. However, only adding the DVFs without adding data consistency can be detrimental to the image reconstruction quality. At \(R_{\text{4D}}<4\), adding DVFs to the images resulted in a lower SSIM, as indicated in \autoref{fig:comparison}. However, at \(R_{\text{4D}}=4\) and in combination with data consistency, increased SSIM, lower EPE, and lower NRMSE was observed by adding DVFs. This could indicate that adding motion information at higher undersampling helps image reconstruction but provides less benefit at lower undersampling factors. This latter aspect could be due to the better conditioning of the inverse problem at higher sampling factors and due to imperfections in the motion estimation model. Currently, we only present the DVFs to the model as generated by a pre-trained network\cite{Terpstra_2020}, which could limit the model performance. Based on previous literature, we foresee that performance may be improved by jointly learning the image reconstruction and DVFs during training\cite{https://doi.org/10.1002/mrm.26206,10.1007/978-3-030-00934-2_53}, improving image registration and image reconstruction performance.

Also, it would be interesting to investigate whether combining the spatial and temporal features by a learnable operator, e.g., convolution or self-attention\cite{NIPS2017_3f5ee243}, would impact, possibly improving the model performance and leading to even shorter MRI acquisitions.
Alternatively, one could optimize the imaging protocol whenever possible by refining the image contrast and reducing scan time by decreasing the number of slices while maintaining the large field of view by slice interpolation. 

This work used XD-GRASP reconstructed 4D-MRI as a ground truth since it demonstrated sufficient accuracy for radiotherapy applications\cite{https://doi.org/10.1002/mrm.28200,PAULSON202072,Mickevicius_2017}. However, this algorithm's regularization over the respiratory phases can introduce errors by overly smoothing the respiratory motion. This could introduce differences in motion amplitude compared to the measured data, and this uncertainty might limit the reconstructed motion quality by deep learning models. 
Using iterative joint image and motion reconstruction as ground truth could be a viable way to improve image quality\cite{https://doi.org/10.1002/mrm.26206} and remove residual artifacts in the ground truth.
When comparing to XD-GRASP we considered a GPU implementation using commodity hardware, which might not be optimal.
Technological developments have accelerate the XD-GRASP algorithm with specialized "Processing-in-memory" hardware\cite{9651604}, curtailing the computational bottleneck for XD-GRASP which enables a speed-up factor of 11, or 90 seconds of processing time.  However, while this is a promising approach, these speed-ups have only been achieved in simulation and such hardware has not been clinically demonstrated.

The models presented in this manuscript have been trained on data obtained from eighteen patients, which is a limited training set size and could limit the performance of the presented models. Large training sets can offer several advantages, such as better performance and improved generalization capabilities. Several steps can be taken to increase the size of our training set. First, more patient data could be acquired, but this process is slow and costly, resulting in limited extra data. Second, digital phantoms could be used to generate 4D-MRI from numerical anatomy\cite{segars_2008}. However, these samples might not be accurate compared to 4D-MRI acquired in-vivo. Future work will investigate the impact of different data augmentation approaches and dataset size. 

MODEST is not the only architecture able to reconstruct 3D+t MRI. Freedman et al. proposed the so-called Dracula framework\cite{FREEDMAN2021209}, consisting of a U-Net reconstructing zero-filled radial 4D-MRI to a high-quality 4D-MRI dataset and a mid-position image. 
Dracula produced 4D-MRI similar to HDTV-MoCo-based 4D-MRI in 28 seconds. 
However, this model was only investigated with a five-minute acquisition. Moreover, the network consisted of approximately 90,000,000 trainable parameters and took 11 days of training. 
Given the number of trainable parameters and their related GPU memory consumption, extending the model from a slice-by-slice reconstruction to a four-dimensional reconstruction is challenging. 
K{\"u}stner et al. proposed CINENet: a complex-valued unrolled U-Net that performs 4D spatio-temporal convolutions to reconstruct cardiac phase-resolved 4D-MRI \cite{Kuestner2020}. They achieve the 4D convolutions by interspersing 3D convolutions with 1D convolutions. CINENet used an approach somewhat similar to ours by decomposing the 4D convolution into lower-dimensional convolution kernels, but we separated the spatial and temporal domains, whereas in CINENet they are interspersed. It is currently unclear whether interspersing or separating the spatial and temporal features would result in better performance, and it may be the object of future investigations. 

MODEST has been specifically constructed to take advantage of the spatio-temporal information in 4D-MRI to obtain high-quality reconstructions.
Interestingly, spatial and temporal information from MRI is relevant in other applications, such as cardiac imaging \cite{Kuestner2020,machado2022deep} or dynamic contrast-enhanced MRI \cite{https://doi.org/10.1002/mrm.24710,Chen2017}. 
Future work could investigate the application of MODEST, retraining the currently used model for these applications.

The availability of fast, accurate, and high-quality 4D-MRI is of particular interest for MRI-guided radiotherapy, where 4D-MRI is used for treatment adaptation of mobile tumors. 
With fast acquisition and reconstruction of 4D-MRI, treatment efficiency and patient comfort can be improved, eliminating the acquisition of a 4D-CT for motion quantification. 
By treating such patients on a hybrid MRI-Linac, motion can quickly be quantified without repositioning the patient.  
Moreover, high-quality 4D-MRI can also be used for high-quality time-resolved imaging\cite{https://doi.org/10.1002/mrm.28200,Kim_2021} and could be helpful for real-time intra-fraction radiation treatment adaptation\cite{KEALL2019228}.

\section{Conclusion}
We proposed a deep learning architecture called MODEST that efficiently reconstructs high-quality 4D-MRI by decomposing the reconstruction into spatial and temporal components. 
This approach yielded superior performance than conventional models such as U-Nets, despite having only 3\% of the trainable parameters. 
We found that high-quality 4D-MRI can be obtained with an MR acquisition of two minutes and 15 seconds of model inference, shortening the time for MRI-guided radiation treatments while improving treatment quality and incorporating accurate motion quantification.
\section{Acknowledgement}
This work is part of the research program HTSM with project number 15354, which is (partly) financed by the Netherlands Organisation for Scientific Research (NWO) and Philips Healthcare. We gratefully acknowledge the support of NVIDIA Corporation with the donation of the Quadro RTX 5000 GPU used for prototyping this research.

\section*{References}
\addcontentsline{toc}{section}{\numberline{}References}
\vspace*{-20mm}





\bibliography{./latex_template_MedPhys_2021}      



\bibliographystyle{./medphy.bst}    


\end{document}

%% file: table.tex
\begin{table}
\caption{\textbf{Quantitative evaluation of the test set for free-breathing undersampled 4D-MRI using 500 spokes}. Best result per metric marked in boldface, results with a star for MODEST indicate a statistically significant improvement compared U-Net (\(p < 0.05\)).}
\resizebox{\linewidth}{!}{%

\begin{tabular}{l|llll}
{} &      SSIM (\(\uparrow\))  &     NRMSE (\(\downarrow\))  &  Sharpness (\(\uparrow\))  &       EPE (\(\downarrow\)) \\
\hline
Zero-filled   &  \(0.689\pm0.019\) &  \(0.673\pm0.07\) &   \(\mathbf{0.104\pm0.006}\) &  \(0.813\pm0.39\) \\
U-Net         &  \(0.871\pm0.032\) &  \(0.824\pm0.16\) &   \(0.030\pm0.005\) &  \(0.326\pm0.20\) \\
MODEST    &  \(\mathbf{0.877\pm0.025}\) &  \(\mathbf{0.383\pm0.11^*}\) &   \(0.043\pm0.007^*\) &  \(\mathbf{0.313\pm0.20}\) \\
Pruned U-Net  &  \(0.801\pm0.036\) &  \(0.920\pm0.11\) &   \(0.036\pm0.006\) &  \(0.512\pm0.24\) 
\end{tabular}
}\label{tab:quant_600_tab}

\end{table}

%% file: latex_template_MedPhys_2021.bbl
\begin{thebibliography}{10}

\bibitem{https://doi.org/10.1118/1.4754658}
J.~C. Park, S.~H. Park, J.~H. Kim, S.~M. Yoon, S.~Y. Song, Z.~Liu, B.~Song,
  K.~Kauweloa, M.~J. Webster, A.~Sandhu, L.~K. Mell, S.~B. Jiang, A.~J. Mundt,
  and W.~Y. Song,
\newblock Liver motion during cone beam computed tomography guided stereotactic
  body radiation therapy,
\newblock Medical Physics {\bf 39}, 6431--6442 (2012).

\bibitem{EKBERG199871}
L.~Ekberg, O.~Holmberg, L.~Wittgren, G.~Bjelkengren, and T.~Landberg,
\newblock What margins should be added to the clinical target volume in
  radiotherapy treatment planning for lung cancer?,
\newblock Radiotherapy and Oncology {\bf 48}, 71--77 (1998).

\bibitem{https://doi.org/10.1002/jmri.22418}
R.~Song, A.~Tipirneni, P.~Johnson, R.~B. Loeffler, and C.~M. Hillenbrand,
\newblock Evaluation of respiratory liver and kidney movements for MRI
  navigator gating,
\newblock Journal of Magnetic Resonance Imaging {\bf 33}, 143--148 (2011).

\bibitem{FENG2009884}
M.~Feng, J.~M. Balter, D.~Normolle, S.~Adusumilli, Y.~Cao, T.~L. Chenevert, and
  E.~Ben-Josef,
\newblock Characterization of Pancreatic Tumor Motion Using Cine MRI:
  Surrogates for Tumor Position Should Be Used With Caution,
\newblock International Journal of Radiation Oncology*Biology*Physics {\bf 74},
  884--891 (2009).

\bibitem{Wood1985MRIA}
M.~L. Wood and R.~M. Henkelman,
\newblock MR image artifacts from periodic motion.,
\newblock Medical physics {\bf 12 2}, 143--51 (1985).

\bibitem{NOTERDAEME2007273}
O.~Noterdaeme, F.~Gleeson, R.~R. Phillips, and M.~Brady,
\newblock Quantification of missing and overlapping data in multiple breath
  hold abdominal imaging,
\newblock European Journal of Radiology {\bf 64}, 273--278 (2007),
\newblock Ultrasound Imaging Special Issue.

\bibitem{OZHASOGLU20021389}
C.~Ozhasoglu and M.~J. Murphy,
\newblock Issues in respiratory motion compensation during external-beam
  radiotherapy,
\newblock International Journal of Radiation Oncology*Biology*Physics {\bf 52},
  1389--1399 (2002).

\bibitem{https://doi.org/10.1118/1.2349696}
P.~J. Keall, G.~S. Mageras, J.~M. Balter, R.~S. Emery, K.~M. Forster, S.~B.
  Jiang, J.~M. Kapatoes, D.~A. Low, M.~J. Murphy, B.~R. Murray, C.~R. Ramsey,
  M.~B. Van~Herk, S.~S. Vedam, J.~W. Wong, and E.~Yorke,
\newblock The management of respiratory motion in radiation oncology report of
  AAPM Task Group 76a),
\newblock Medical Physics {\bf 33}, 3874--3900 (2006).

\bibitem{BUSSELS200369}
B.~Bussels, L.~Goethals, M.~Feron, D.~Bielen, S.~Dymarkowski, P.~Suetens, and
  K.~Haustermans,
\newblock Respiration-induced movement of the upper abdominal organs: a pitfall
  for the three-dimensional conformal radiation treatment of pancreatic cancer,
\newblock Radiotherapy and Oncology {\bf 68}, 69--74 (2003).

\bibitem{Vedam_2002}
S.~S. Vedam, P.~J. Keall, V.~R. Kini, H.~Mostafavi, H.~P. Shukla, and R.~Mohan,
\newblock Acquiring a four-dimensional computed tomography dataset using an
  external respiratory signal,
\newblock Physics in Medicine and Biology {\bf 48}, 45--62 (2002).

\bibitem{https://doi.org/10.1002/mrm.25753}
Z.~Deng, J.~Pang, W.~Yang, Y.~Yue, B.~Sharif, R.~Tuli, D.~Li, B.~Fraass, and
  Z.~Fan,
\newblock Four-dimensional MRI using three-dimensional radial sampling with
  respiratory self-gating to characterize temporal phase-resolved respiratory
  motion in the abdomen,
\newblock Magnetic Resonance in Medicine {\bf 75}, 1574--1585 (2016).

\bibitem{RIETZEL2006287}
E.~Rietzel, A.~K. Liu, K.~P. Doppke, J.~A. Wolfgang, A.~B. Chen, G.~T. Chen,
  and N.~C. Choi,
\newblock Design of 4D treatment planning target volumes,
\newblock International Journal of Radiation Oncology*Biology*Physics {\bf 66},
  287--295 (2006).

\bibitem{SENTKER2020229}
T.~Sentker, V.~Schmidt, A.-K. Ozga, C.~Petersen, F.~Madesta, C.~Hofmann,
  R.~Werner, and T.~Gauer,
\newblock 4D CT image artifacts affect local control in SBRT of lung and liver
  metastases,
\newblock Radiotherapy and Oncology {\bf 148}, 229--234 (2020).

\bibitem{https://doi.org/10.1118/1.2717404}
Y.~D. Mutaf, J.~A. Antolak, and D.~H. Brinkmann,
\newblock The impact of temporal inaccuracies on 4DCT image quality,
\newblock Medical Physics {\bf 34}, 1615--1622 (2007).

\bibitem{Raaymakers_2009}
B.~W. Raaymakers, J.~J.~W. Lagendijk, J.~Overweg, J.~G.~M. Kok, A.~J.~E.
  Raaijmakers, E.~M. Kerkhof, R.~W. van~der Put, I.~Meijsing, S.~P.~M. Crijns,
  F.~Benedosso, M.~van Vulpen, C.~H.~W. de~Graaff, J.~Allen, and K.~J. Brown,
\newblock Integrating a 1.5 T {MRI} scanner with a 6 {MV} accelerator: proof of
  concept,
\newblock Physics in Medicine and Biology {\bf 54}, N229--N237 (2009).

\bibitem{MUTIC2014196}
S.~Mutic and J.~F. Dempsey,
\newblock The ViewRay System: Magnetic Resonance–Guided and Controlled
  Radiotherapy,
\newblock Seminars in Radiation Oncology {\bf 24}, 196--199 (2014),
\newblock Magnetic Resonance Imaging in Radiation Oncology.

\bibitem{https://doi.org/10.1002/mp.15217}
M.~L. Terpstra, M.~Maspero, T.~Bruijnen, J.~J. Verhoeff, J.~J. Lagendijk, and
  C.~A. van~den Berg,
\newblock Real-time 3D motion estimation from undersampled MRI using
  multi-resolution neural networks,
\newblock Medical Physics {\bf 48}, 6597--6613 (2021).

\bibitem{Huttinga_2020}
N.~R.~F. Huttinga, C.~A.~T. van~den Berg, P.~R. Luijten, and A.~Sbrizzi,
\newblock {MR}-{MOTUS}: model-based non-rigid motion estimation for {MR}-guided
  radiotherapy using a reference image and minimal k-space data,
\newblock Physics in Medicine {\&} Biology {\bf 65}, 015004 (2020).

\bibitem{Glitzner_2015}
M.~Glitzner, B.~D. de~Senneville, J.~J.~W. Lagendijk, B.~W. Raaymakers, and
  S.~P.~M. Crijns,
\newblock On-line 3Dmotion estimation using low resolution {MRI},
\newblock Physics in Medicine and Biology {\bf 60}, N301--N310 (2015).

\bibitem{MENTEN2017139}
M.~J. Menten, A.~Wetscherek, and M.~F. Fast,
\newblock MRI-guided lung SBRT: Present and future developments,
\newblock Physica Medica {\bf 44}, 139--149 (2017).

\bibitem{https://doi.org/10.1118/1.4927252}
C.~Paganelli, P.~Summers, M.~Bellomi, G.~Baroni, and M.~Riboldi,
\newblock Liver 4DMRI: A retrospective image-based sorting method,
\newblock Medical Physics {\bf 42}, 4814--4821 (2015).

\bibitem{KEALL2019228}
P.~Keall, P.~Poulsen, and J.~T. Booth,
\newblock See, Think, and Act: Real-Time Adaptive Radiotherapy,
\newblock Seminars in Radiation Oncology {\bf 29}, 228--235 (2019),
\newblock Adaptive Radiotherapy and Automation.

\bibitem{PAULSON202072}
E.~S. Paulson, E.~Ahunbay, X.~Chen, N.~J. Mickevicius, G.-P. Chen, C.~Schultz,
  B.~Erickson, M.~Straza, W.~A. Hall, and X.~A. Li,
\newblock 4D-MRI driven MR-guided online adaptive radiotherapy for abdominal
  stereotactic body radiation therapy on a high field MR-Linac: Implementation
  and initial clinical experience,
\newblock Clinical and Translational Radiation Oncology {\bf 23}, 72--79
  (2020).

\bibitem{Mickevicius_2017}
N.~J. Mickevicius and E.~S. Paulson,
\newblock Investigation of undersampling and reconstruction algorithm
  dependence on respiratory correlated 4D-{MRI} for online {MR}-guided
  radiation therapy,
\newblock Physics in Medicine and Biology {\bf 62}, 2910--2921 (2017).

\bibitem{6773024}
C.~E. Shannon,
\newblock A mathematical theory of communication,
\newblock The Bell System Technical Journal {\bf 27}, 379--423 (1948).

\bibitem{sense_99}
K.~P. Pruessmann, M.~Weiger, M.~B. Scheidegger, and P.~Boesiger,
\newblock SENSE: Sensitivity encoding for fast MRI,
\newblock Magnetic Resonance in Medicine {\bf 42}, 952--962 (1999).

\bibitem{https://doi.org/10.1002/mrm.10171}
M.~A. Griswold, P.~M. Jakob, R.~M. Heidemann, M.~Nittka, V.~Jellus, J.~Wang,
  B.~Kiefer, and A.~Haase,
\newblock Generalized autocalibrating partially parallel acquisitions (GRAPPA),
\newblock Magnetic Resonance in Medicine {\bf 47}, 1202--1210 (2002).

\bibitem{sms_01}
D.~J. Larkman, J.~V. Hajnal, A.~H. Herlihy, G.~A. Coutts, I.~R. Young, and
  G.~Ehnholm,
\newblock Use of multicoil arrays for separation of signal from multiple slices
  simultaneously excited,
\newblock Journal of Magnetic Resonance Imaging {\bf 13}, 313--317 (2001).

\bibitem{Keijnemans_2021}
K.~Keijnemans, P.~T.~S. Borman, A.~L. H. M.~W. van Lier, J.~J.~C. Verhoeff,
  B.~W. Raaymakers, and M.~F. Fast,
\newblock Simultaneous multi-slice accelerated 4D-{MRI} for radiotherapy
  guidance,
\newblock Physics in Medicine \& Biology {\bf 66}, 095014 (2021).

\bibitem{https://doi.org/10.1002/mp.15802}
K.~Keijnemans, P.~T.~S. Borman, P.~Uijtewaal, P.~L. Woodhead, B.~W. Raaymakers,
  and M.~F. Fast,
\newblock A hybrid 2D/4D-MRI methodology using simultaneous multislice imaging
  for radiotherapy guidance,
\newblock Medical Physics {\bf n/a}.

\bibitem{https://doi.org/10.1002/mrm.21391}
M.~Lustig, D.~Donoho, and J.~M. Pauly,
\newblock Sparse MRI: The application of compressed sensing for rapid MR
  imaging,
\newblock Magnetic Resonance in Medicine {\bf 58}, 1182--1195 (2007).

\bibitem{https://doi.org/10.1002/mrm.25665}
L.~Feng, L.~Axel, H.~Chandarana, K.~T. Block, D.~K. Sodickson, and R.~Otazo,
\newblock XD-GRASP: Golden-angle radial MRI with reconstruction of extra
  motion-state dimensions using compressed sensing,
\newblock Magnetic Resonance in Medicine {\bf 75}, 775--788 (2016).

\bibitem{https://doi.org/10.1002/mrm.26206}
C.~M. Rank, T.~Heußer, M.~T.~A. Buzan, A.~Wetscherek, M.~T. Freitag,
  J.~Dinkel, and M.~Kachelrieß,
\newblock 4D respiratory motion-compensated image reconstruction of
  free-breathing radial MR data with very high undersampling,
\newblock Magnetic Resonance in Medicine {\bf 77}, 1170--1183 (2017).

\bibitem{https://doi.org/10.1002/mrm.26977}
K.~Hammernik, T.~Klatzer, E.~Kobler, M.~P. Recht, D.~K. Sodickson, T.~Pock, and
  F.~Knoll,
\newblock Learning a variational network for reconstruction of accelerated MRI
  data,
\newblock Magnetic Resonance in Medicine {\bf 79}, 3055--3071 (2018).

\bibitem{8067520}
J.~Schlemper, J.~Caballero, J.~V. Hajnal, A.~N. Price, and D.~Rueckert,
\newblock A Deep Cascade of Convolutional Neural Networks for Dynamic MR Image
  Reconstruction,
\newblock IEEE Transactions on Medical Imaging {\bf 37}, 491--503 (2018).

\bibitem{10.1007/978-3-030-59713-9_7}
A.~Sriram, J.~Zbontar, T.~Murrell, A.~Defazio, C.~L. Zitnick, N.~Yakubova,
  F.~Knoll, and P.~Johnson,
\newblock End-to-End Variational Networks for Accelerated MRI Reconstruction,
\newblock in {\em Medical Image Computing and Computer Assisted Intervention --
  MICCAI 2020}, edited by A.~L. Martel, P.~Abolmaesumi, D.~Stoyanov, D.~Mateus,
  M.~A. Zuluaga, S.~K. Zhou, D.~Racoceanu, and L.~Joskowicz, pages 64--73,
  Cham, 2020, MICCAI, Springer International Publishing.

\bibitem{https://doi.org/10.1002/mrm.27706}
S.~Biswas, H.~K. Aggarwal, and M.~Jacob,
\newblock Dynamic MRI using model-based deep learning and SToRM priors:
  MoDL-SToRM,
\newblock Magnetic Resonance in Medicine {\bf 82}, 485--494 (2019).

\bibitem{9048706}
C.~Zhang, S.~A. Hossein~Hosseini, S.~Moeller, S.~Weingärtner, K.~Ugurbil, and
  M.~Akcakaya,
\newblock Scan-Specific Residual Convolutional Neural Networks for Fast MRI
  Using Residual RAKI,
\newblock in {\em 2019 53rd Asilomar Conference on Signals, Systems, and
  Computers}, pages 1476--1480, IEEE, 2019.

\bibitem{machado2022deep}
I.~P. Machado, E.~Puyol-Antón, K.~Hammernik, G.~Cruz, D.~Ugurlu, I.~Olakorede,
  I.~Oksuz, B.~Ruijsink, M.~Castelo-Branco, A.~A. Young, C.~Prieto, J.~A.
  Schnabel, and A.~P. King,
\newblock A Deep Learning-based Integrated Framework for Quality-aware
  Undersampled Cine Cardiac MRI Reconstruction and Analysis,
\newblock \url{https://arxiv.org/abs/2205.01673}, 2022.

\bibitem{FREEDMAN2021209}
J.~N. Freedman, O.~J. Gurney-Champion, S.~Nill, A.-M. Shiarli, H.~E.
  Bainbridge, H.~C. Mandeville, D.-M. Koh, F.~McDonald, M.~Kachelrieß,
  U.~Oelfke, and A.~Wetscherek,
\newblock Rapid 4D-MRI reconstruction using a deep radial convolutional neural
  network: Dracula,
\newblock Radiotherapy and Oncology {\bf 159}, 209--217 (2021).

\bibitem{Kuestner2020}
T.~K{\"u}stner, N.~Fuin, K.~Hammernik, A.~Bustin, H.~Qi, R.~Hajhosseiny, P.~G.
  Masci, R.~Neji, D.~Rueckert, R.~M. Botnar, and C.~Prieto,
\newblock CINENet: deep learning-based 3D cardiac CINE MRI reconstruction with
  multi-coil complex-valued 4D spatio-temporal convolutions,
\newblock Scientific Reports {\bf 10}, 13710 (2020).

\bibitem{8425639}
C.~Qin, J.~Schlemper, J.~Caballero, A.~N. Price, J.~V. Hajnal, and D.~Rueckert,
\newblock Convolutional Recurrent Neural Networks for Dynamic MR Image
  Reconstruction,
\newblock IEEE Transactions on Medical Imaging {\bf 38}, 280--290 (2019).

\bibitem{8793147}
A.~Kofler, M.~Dewey, T.~Schaeffter, C.~Wald, and C.~Kolbitsch,
\newblock Spatio-Temporal Deep Learning-Based Undersampling Artefact Reduction
  for 2D Radial Cine MRI With Limited Training Data,
\newblock IEEE Transactions on Medical Imaging {\bf 39}, 703--717 (2020).

\bibitem{https://doi.org/10.1002/mrm.25858}
T.~Zhang, J.~Y. Cheng, Y.~Chen, D.~G. Nishimura, J.~M. Pauly, and S.~S.
  Vasanawala,
\newblock {Robust self-navigated body MRI using dense coil arrays},
\newblock {Magnetic Resonance in Medicine} {\bf 76}, 197--205 (2016).

\bibitem{stemkens_hybrid}
B.~Stemkens, R.~H. Tijssen, B.~D. {de Senneville}, H.~D. Heerkens, M.~{van
  Vulpen}, J.~J. Lagendijk, and C.~A. {van den Berg},
\newblock {Optimizing 4-Dimensional Magnetic Resonance Imaging Data Sampling
  for Respiratory Motion Analysis of Pancreatic Tumors},
\newblock International Journal of Radiation Oncology*Biology*Physics {\bf 91},
  571 -- 578 (2015).

\bibitem{1166689}
J.~A. {Fessler} and B.~P. {Sutton},
\newblock {Nonuniform fast Fourier transforms using min-max interpolation},
\newblock IEEE Transactions on Signal Processing {\bf 51}, 560--574 (2003).

\bibitem{knoll2014gpunufft}
F.~Knoll, A.~Schwarzl, C.~Diwoky, and D.~K. Sodickson,
\newblock gpuNUFFT-an open source GPU library for 3D regridding with direct
  Matlab interface,
\newblock page 4297, Proceedings of the 22nd annual meeting of ISMRM, Milan,
  Italy, 2014.

\bibitem{https://doi.org/10.1002/mrm.24751}
M.~Uecker, P.~Lai, M.~J. Murphy, P.~Virtue, M.~Elad, J.~M. Pauly, S.~S.
  Vasanawala, and M.~Lustig,
\newblock ESPIRiT—an eigenvalue approach to autocalibrating parallel MRI:
  Where SENSE meets GRAPPA,
\newblock Magnetic Resonance in Medicine {\bf 71}, 990--1001 (2014).

\bibitem{Lustig2007}
M.~Lustig, D.~Donoho, and J.~M. Pauly,
\newblock {Sparse MRI: The application of compressed sensing for rapid MR
  imaging},
\newblock Magnetic Resonance in Medicine {\bf 58}, 1182--1195 (2007).

\bibitem{Terpstra_2020}
M.~L. Terpstra, M.~Maspero, F.~d'Agata, B.~Stemkens, M.~P.~W. Intven, J.~J.~W.
  Lagendijk, C.~A.~T. van~den Berg, and R.~H.~N. Tijssen,
\newblock Deep learning-based image reconstruction and motion estimation from
  undersampled radial k-space for real-time {MRI}-guided radiotherapy,
\newblock Physics in Medicine {\&} Biology {\bf 65}, 155015 (2020).

\bibitem{8297024}
P.~Virtue, S.~X. Yu, and M.~Lustig,
\newblock Better than real: Complex-valued neural nets for MRI fingerprinting,
\newblock in {\em 2017 IEEE International Conference on Image Processing
  (ICIP)}, pages 3953--3957, IEEE, 2017.

\bibitem{10.1007/978-3-030-12029-0_40}
E.~Kerfoot, J.~Clough, I.~Oksuz, J.~Lee, A.~P. King, and J.~A. Schnabel,
\newblock Left-Ventricle Quantification Using Residual U-Net,
\newblock in {\em Statistical Atlases and Computational Models of the Heart.
  Atrial Segmentation and LV Quantification Challenges}, edited by M.~Pop,
  M.~Sermesant, J.~Zhao, S.~Li, K.~McLeod, A.~Young, K.~Rhode, and T.~Mansi,
  pages 371--380, Cham, 2019, Springer, Springer International Publishing.

\bibitem{the_monai_consortium_2020_4323059}
T.~M. Consortium,
\newblock Project MONAI,
\newblock online, 2020.

\bibitem{muckley:20:tah}
M.~J. Muckley, R.~Stern, T.~Murrell, and F.~Knoll,
\newblock {TorchKbNufft}: A High-Level, Hardware-Agnostic Non-Uniform Fast
  {Fourier} Transform,
\newblock in {\em ISMRM Workshop on Data Sampling \& Image Reconstruction},
  2020,
\newblock Source code available at
  \url{https://github.com/mmuckley/torchkbnufft}.

\bibitem{perploss}
M.~L. Terpstra, M.~Maspero, A.~Sbrizzi, and C.~A. van~den Berg,
\newblock \(\perp\)-loss: a symmetric loss function for magnetic resonance
  imaging reconstruction and image registration with deep learning,
\newblock Medical Image Analysis {\bf 48}, 6597--6613 (2022).

\bibitem{DBLP:journals/corr/JinYFY16}
X.~Jin, X.~Yuan, J.~Feng, and S.~Yan,
\newblock Training Skinny Deep Neural Networks with Iterative Hard Thresholding
  Methods,
\newblock CoRR {\bf abs/1607.05423} (2016).

\bibitem{https://doi.org/10.1002/mp.15514}
D.~Gourdeau, S.~Duchesne, and L.~Archambault,
\newblock On the proper use of structural similarity for the robust evaluation
  of medical image synthesis models,
\newblock Medical Physics {\bf 49}, 2462--2474 (2022).

\bibitem{VANDELINDT2018875}
T.~{van de Lindt}, J.-J. Sonke, M.~Nowee, E.~Jansen, V.~{van Pelt}, U.~{van der
  Heide}, and M.~Fast,
\newblock A Self-Sorting Coronal 4D-MRI Method for Daily Image Guidance of
  Liver Lesions on an MR-LINAC,
\newblock International Journal of Radiation Oncology*Biology*Physics {\bf
  102}, 875--884 (2018),
\newblock Imaging in Radiation Oncology.

\bibitem{903548}
J.~Pech-Pacheco, G.~Cristobal, J.~Chamorro-Martinez, and J.~Fernandez-Valdivia,
\newblock {Diatom autofocusing in brightfield microscopy: a comparative study},
\newblock in {\em Proceedings 15th International Conference on Pattern
  Recognition. ICPR-2000}, volume~3, pages 314--317, IEEE, IEEE Comput. Soc,
  2000.

\bibitem{6796174}
S.~J. Nowlan and G.~E. Hinton,
\newblock Simplifying Neural Networks by Soft Weight-Sharing,
\newblock Neural Computation {\bf 4}, 473--493 (1992).

\bibitem{10.1093/bib/bbx044}
R.~Miotto, F.~Wang, S.~Wang, X.~Jiang, and J.~T. Dudley,
\newblock {Deep learning for healthcare: review, opportunities and challenges},
\newblock Briefings in Bioinformatics {\bf 19}, 1236--1246 (2017).

\bibitem{9703109}
Y.~Chen, C.-B. Schönlieb, P.~Liò, T.~Leiner, P.~L. Dragotti, G.~Wang,
  D.~Rueckert, D.~Firmin, and G.~Yang,
\newblock AI-Based Reconstruction for Fast MRI—A Systematic Review and
  Meta-Analysis,
\newblock Proceedings of the IEEE {\bf 110}, 224--245 (2022).

\bibitem{10.1007/978-3-030-00934-2_53}
C.~Qin, W.~Bai, J.~Schlemper, S.~E. Petersen, S.~K. Piechnik, S.~Neubauer, and
  D.~Rueckert,
\newblock {Joint Learning of Motion Estimation and Segmentation for Cardiac MR
  Image Sequences},
\newblock in {\em {Medical Image Computing and Computer Assisted Intervention
  -- MICCAI 2018}}, edited by A.~F. Frangi, J.~A. Schnabel, C.~Davatzikos,
  C.~Alberola-L{\'o}pez, and G.~Fichtinger, pages 472--480, Cham, 2018, MICCAI,
  Springer International Publishing.

\bibitem{NIPS2017_3f5ee243}
A.~Vaswani, N.~Shazeer, N.~Parmar, J.~Uszkoreit, L.~Jones, A.~N. Gomez, L.~u.
  Kaiser, and I.~Polosukhin,
\newblock Attention is All you Need,
\newblock in {\em Advances in Neural Information Processing Systems}, edited by
  I.~Guyon, U.~V. Luxburg, S.~Bengio, H.~Wallach, R.~Fergus, S.~Vishwanathan,
  and R.~Garnett, volume~30, NeurIPS, Curran Associates, Inc., 2017.

\bibitem{https://doi.org/10.1002/mrm.28200}
L.~Feng, N.~Tyagi, and R.~Otazo,
\newblock MRSIGMA: Magnetic Resonance SIGnature MAtching for real-time
  volumetric imaging,
\newblock Magnetic Resonance in Medicine {\bf 84}, 1280--1292 (2020).

\bibitem{9651604}
M.~Barbone, A.~Wetscherek, T.~Yung, U.~Oelfke, W.~Luk, and G.~Gaydadjiev,
\newblock Efficient Online 4D Magnetic Resonance Imaging,
\newblock in {\em 2021 IEEE 33rd International Symposium on Computer
  Architecture and High Performance Computing (SBAC-PAD)}, pages 177--187,
  IEEE, 2021.

\bibitem{segars_2008}
W.~P. Segars, M.~Mahesh, T.~J. Beck, E.~C. Frey, and B.~M.~W. Tsui,
\newblock {Realistic CT simulation using the 4D XCAT phantom},
\newblock Medical Physics {\bf 35}, 3800--3808 (2008).

\bibitem{https://doi.org/10.1002/mrm.24710}
R.~M. Lebel, J.~Jones, J.-C. Ferre, M.~Law, and K.~S. Nayak,
\newblock Highly accelerated dynamic contrast enhanced imaging,
\newblock Magnetic Resonance in Medicine {\bf 71}, 635--644 (2014).

\bibitem{Chen2017}
J.~Chen, S.~Liu, and M.~Huang,
\newblock Low-Rank and Sparse Decomposition Model for Accelerating Dynamic MRI
  Reconstruction,
\newblock Journal of Healthcare Engineering {\bf 2017}, 9856058 (2017).

\bibitem{Kim_2021}
N.~Kim, K.~R. Tringale, C.~Crane, N.~Tyagi, and R.~Otazo,
\newblock {MR} {SIGnature} {MAtching} ({MRSIGMA}) with retrospective
  self-evaluation for real-time volumetric motion imaging,
\newblock Physics in Medicine \& Biology {\bf 66}, 215009 (2021).

\end{thebibliography}
